\documentclass{article}
\usepackage{amsmath}
\usepackage{amsfonts}
\usepackage{amssymb}
\usepackage{graphicx}

\input{tcilatex}

\begin{document}

\title{Relative Entropy and Inductive Inference\thanks{%
Presented at MaxEnt 2003, the 23th International Workshop on Bayesian
Inference and Maximum Entropy Methods (August 3-8, Jackson Hole, WY, USA).}}
\author{Ariel Caticha \\
{\small Department of Physics, University at Albany-SUNY, }\\
{\small Albany, NY 12222, USA.\thanks{%
E-mail: ariel@albany.edu}}}
\date{}
\maketitle

\begin{abstract}
We discuss how the method of maximum entropy, MaxEnt, can be extended beyond
its original scope, as a rule to assign a probability distribution, to a
full-fledged method for inductive inference. The main concept is the
(relative) entropy $S[p|q]$ which is designed as a tool to update from a
prior probability distribution $q$ to a posterior probability distribution $%
p $ when new information in the form of a constraint becomes available. The
extended method goes beyond the mere selection of a single posterior $p$,
but also addresses the question of how much less probable other
distributions might be. Our approach clarifies how the entropy $S[p|q]$ is
used while avoiding the question of its meaning. Ultimately, entropy is a
tool for induction which needs no interpretation. Finally, being a tool for
generalization from special examples, we ask whether the functional form of
the entropy depends on the choice of the examples and we find that it does.
The conclusion is that there is no single general theory of inductive
inference and that alternative expressions for the entropy are possible.
\end{abstract}

\section{Introduction}

The method of maximum entropy, MaxEnt, as conceived by Jaynes \cite{Jaynes57}%
, is a method to assign probabilities on the basis of partial information of
a certain kind. The type of information in question is called testable
information and consists in the specification of the family of acceptable
distributions. The information is \textquotedblleft
testable\textquotedblright\ in the sense that one should be able to test
whether any candidate distribution belongs or not to the family.

The purpose of this paper is to discuss how MaxEnt can be extended beyond
its original scope, as a rule to assign a probability distribution, to a
full-fledged method for inductive inference. To distinguish it from MaxEnt
the extended method will henceforth be abbreviated as ME. \cite{footnote1}

The general problem of inductive inference is to update from a prior
probability distribution to a posterior distribution when new information
becomes available. The challenge is to develop updating methods that are
systematic and objective. Two methods have been found which are of very
broad applicability: one is based on Bayes' theorem and the other is ME. The
choice between these two updating methods is dictated by the nature of the
information being processed.

When we want to update our beliefs about the values of certain quantities $%
\theta $ on the basis of information about the observed values of other
quantities $x$ -- the data -- and of the known relation between them -- the
conditional distribution $p(x|\theta )$ -- we must use Bayes' theorem. If
the prior beliefs are given by $p(\theta )$, the updated or posterior
distribution is $p(\theta |x)\propto p(\theta )p(x|\theta )$. Being a
consequence of the product rule for probabilities, the Bayesian method of
updating is limited to situations where it makes sense to define the joint
probability of $x$ and $\theta $. The ME method, on the other hand, is
designed for updating from a prior probability distribution to a posterior
distribution when the information to be processed is testable information, 
\emph{i.e.}, it takes the form of constraints on the family of acceptable
posterior distributions \cite{footnote2}. In general it makes no sense to
process testable information using Bayes' theorem, and conversely, it makes
no sense to process data using ME. However, in those special cases when the
same piece of information can be both interpreted as data and as a
constraint then both methods can be used and they agree.

There are several justifications for the MaxEnt method. The earliest one,
the multiplicity argument, dates back to Boltzmann and Gibbs and is purely
probabilistic. One counts the number of microstates that are compatible with
each macrostate. Assuming that all microstates are equally likely, the most
probable macrostate is that with the largest number of microstates.

The next justification of the MaxEnt method followed from interpreting
entropy, through the Shannon axioms, as a measure of the \textquotedblleft
amount of uncertainty\textquotedblright\ or of the \textquotedblleft amount
of information that is missing\textquotedblright\ in a probability
distribution \cite{Shannon48, Jaynes57}. One limitation of this approach is
that the Shannon axioms refer to probabilities of discrete variables; for
continuous variables the entropy is not defined. But more serious objections
can be raised, namely, even if we grant that the Shannon axioms do lead to a
reasonable expression for the entropy, to what extent do we believe the
axioms themselves? Shannon's third axiom, the grouping property, is indeed
very reasonable, but is it necessary? Is entropy the only consistent measure
of uncertainty or of information? What is wrong with, say, the standard
deviation? Indeed, there even exist examples in which the entropy does not
seem to reflect one's intuitive notion of information (see \emph{e.g.}, \cite%
{Uffink95}). Other entropies, justified by a\ different choice of axioms,
were subsequently introduced \cite{Renyi61}-\cite{Tsallis88}.

From our point of view the real limitation is that Shannon was not concerned
with inductive inference. He was analyzing the capacity of communication
channels. Shannon's entropy makes no reference to prior distributions.
Indeed, MaxEnt, was conceived by Jaynes as a method of inference, on the
basis of testable information and an underlying physical measure. He never
meant to update from one probability distribution to another, and there was
no induction in the sense that no generalization from special cases was
involved.

Considerations such as these motivated several attempts to develop the ME
method directly as a method for updating probabilities without invoking
questionable measures of uncertainty; prominent among these are the works by
Shore and Johnson \cite{ShoreJohnson80}, Skilling \cite{Skilling88}-\cite%
{Skilling90}, and Csiszar \cite{Csiszar91}.

The important contribution by Shore and Johnson was the realization that one
could axiomatize the updating method itself rather than the information
measure; they propose four axioms and show that the relative entropy is the
unique solution. Their axioms are justified on the basis of a fundamental
principle of consistency -- if a problem can be solved in more than one way
the results should agree -- but the axioms themselves and other assumptions
they make have raised some objections \cite{Karbelkar86, Uffink95}. Despite
such criticism the enormous influence of their pioneering papers is evident.

Skilling derives the maximum entropy method from axioms that are clearly
inspired by those of Shore and Johnson but his approach is different in
several important aspects. First, he broadens the subject matter beyond
probabilities to the determination of other positive-valued functions such
as, for example, intensities in an image. Whether this is a step forward is
debatable. For example, the broader scope of the method makes it riskier and
the justification of the axioms becomes a more delicate matter. Also, the
extension beyond probabilities to positive-valued functions does not
necessarily represent a wider range of applicability. After all,
probabilities already provide us with the tools required for reasoning under
uncertainty, and once we can manipulate them, the reconstruction of all
sorts of other functions, including positive-valued ones, should be tackled
using Bayes' theorem.

Second, and from our point of view, most important: Skilling spells out the
strategy one should follow to construct a general theory based on the
analysis of a few simple examples. This is a remarkable achievement for it
constitutes nothing less than a systematic quantitative method for
induction, for generalizing from special cases \cite{Skilling89}. However,
Skilling does not explore the possibility of using his method for the
purpose of updating probabilities; clearly this was not his immediate goal.

The primary goal of this paper (sections 2 and 3) is to apply Skilling's
method of induction to Shore and Johnson's problem of updating probabilities
and, in the process, hopefully overcome at least some of the objections that
can be raised against either.

The procedure we follow differs in one remarkable way from the manner that
has in the past been followed in setting up physical theories. Normally one
starts by establishing a mathematical formalism, setting up a set of
equations, and then one tries to append an interpretation to it. This is a
very difficult problem; historically it has affected not only statistics and
statistical physics -- what is the meaning of probabilities and of entropy
-- but also quantum theory -- what is the meaning of wave functions and
amplitudes. The issue of whether the proposed interpretation is unique, or
even whether it is allowed, always remains a legitimate objection and a
point of controversy.

Here we proceed in the opposite order, we first decide what we are talking
about and what we want to accomplish, and only afterwards we design the
appropriate mathematical formalism. The advantage is that the issue of
meaning never arises. The preeminent example of this approach is Cox's
algebra of probable inference which clarified the meaning and use of the
notion of probability: after Cox it was no longer possible to raise doubts
about the legitimacy of the degree of belief interpretation. A second
example is special relativity: the actual physical significance of the $x$
and $t$ appearing in the mathematical formalism of Lorentz and Poincare was
a matter of controversy until Einstein settled the issue by deriving the
formalism, \emph{i.e.}, the Lorentz transformations, from more basic
principles. Yet a third example relevant to quantum theory is given in \cite%
{Caticha98}. In this paper we explore a fourth example: the concept of
relative entropy is introduced as a tool for reasoning which, in the special
case of uniform priors, reduces to the usual entropy. There is no need for
an interpretation in terms of heat, multiplicity of states, disorder, or
uncertainty, or even in terms of an amount of information. Perhaps this is
the explanation of why the search for the meaning of entropy has turned out
to be so elusive: ultimately, \emph{entropy needs no interpretation}. We do
not need to know what `entropy' means, we only need to know how to use it.

There is a second function that the ME method must perform in order to
succeed as a method of inductive inference: once we have decided that the
distribution of maximum entropy is to be preferred over all others we must
address the question of how reliable our choice is. In other words, to what
extent do we rule out all those distributions with entropies less than the
maximum. This matter is addressed in Section 4 following the treatment in 
\cite{Caticha00}. In Section 5 we collect miscellaneous remarks on the
choice and nature of the prior distribution, on using entropy as a measure
of amount of information, on choosing constraints, and on the choices of
axioms and how they are justified by other authors.

Since information of different kinds, data or testable information, is meant
to be processed using different methods, it is quite clear that there is no
universal rule for processing information, there is no universal theory of
inductive inference. But we can still ask whether there exists a theory of
induction that is sufficiently general for processing all testable
information. In other words, does the entropy depend on the particular
special examples from which one is generalizing? Is there a unique entropy?
In Section 6 we find that a different choice of special examples does,
indeed, lead to a different functional form for the entropy: there is no
general theory for processing testable information. A summary of our
conclusions and some final comments are collected in section 7.

\section{Entropy as a tool for induction}

Consider a variable $x$ in a space $X$; $x$ could be a discrete or a
continuous variable, in one or several dimensions. It might, for example,
represent the possible microstates of a physical system: $x$ can be a point
in phase space, or an appropriate set of quantum numbers. Our uncertainty
about $x$ is described by a probability distribution $q(x)$. Our goal is to
update from the prior distribution $q(x)$ to a posterior distribution $p(x)$
when new information in the form of a constraint becomes available. The
constraints can, but need not, be linear. The question is what distribution $%
p(x)$ should we select?

To select the posterior one could proceed by attempting to place all
distributions in increasing \emph{order of preference}. Irrespective of what
it is that makes one distribution preferable over another it is clear any
such ranking must be transitive: if distribution $p_{1}$ is preferred over
distribution $p_{2}$, and $p_{2}$ is preferred over $p_{3}$, then $p_{1}$ is
preferred over $p_{3}$. Such transitive rankings are implemented by
assigning to each $p(x)$ a real number $S[p]$ in such a way that if $p_{1}$
is preferred over $p_{2}$, then $S[p_{1}]>S[p_{2}]$. The selected $p$ will
be that which maximizes the functional $S[p]$ which will be called the
entropy of $p$. Thus the ME method involves entropies which are real numbers
and that are meant to be maximized. These are features imposed by design;
they are dictated by the function that the ME method is supposed to perform .

Next, to define the ranking scheme, we must decide on the functional form of 
$S[p]$. \emph{The purpose of the method is to do induction.} We want to
generalize from those special cases where we know what the preferred
distribution should be to the much larger number of cases where we do not.
Thus, in order to achieve its purpose, $S[p]$ will have to be of very
general applicability; we will initially assume that \emph{the same }$S[p]$ 
\emph{applies to all cases}. There is no justification for this generality
beyond the usual pragmatic justification of induction: we must risk making
wrong generalizations in order to avoid the paralysis of not generalizing at
all.

The fundamental inductive principle is the seemingly trivial statement that 
\emph{`If a general theory exists, then it must apply to special cases' \cite%
{Skilling89}.} But the triviality is deceptive: the full power of the
principle becomes clear once we realize that if there exists a special case
where the preferred distribution happens to be known, then this knowledge
can be used to constrain the form of $S[p]$ and, further, if a sufficient
number of constraining examples happens to be known, then $S[p]$ can be
determined completely. Of course, it is quite possible that there be too
many such constraints, that there is no $S[p]$ satisfying them all. One
would then be forced to conclude that there is no general theory. In such a
situation the best one can do is produce theories of inductive inference
that are not completely general but that can still be useful if their range
of applicability is sufficiently wide.

The presumably \textquotedblleft known\textquotedblright\ special cases,
called the \textquotedblleft axioms\textquotedblright\ of the theory, play a
crucial role: their choice defines which general theory is being
constructed. In our case, we want to design a theory for updating
probability distributions. The axioms below are chosen to reflect the
conviction that one should not change one's mind frivolously, that the only
aspects of one's beliefs that should be updated are those for which new
evidence has been supplied. Our approach is remarkably cautious; the axioms
do not tell us what and how to update, they merely tell us what not to
update.

Degrees of belief, probabilities, are said to be subjective; two different
individuals might not share the same beliefs and could conceivably assign
probabilities differently. But subjectivity does not mean arbitrariness. The
reason subjective probabilities are introduced in the first place is to
bring objectivity into our reasoning: the subjectivity of probabilities does
not extend to allow us to assign some probabilities now and later revise
them unless forced by new information that has in the meantime become
available \cite{Diaconis82}. This \textquotedblleft innocent until proven
guilty\textquotedblright\ attitude is designed to maximize objectivity:
There are many ways to change but only one to remain the same. It is also a
recognition of the high value we place on the prior probabilities which
codify information that was laboriously collected and processed in the past.

The three axioms and their consequences are listed below. The proofs are
given in the next section.

\textbf{Axiom 1: Locality}. \emph{Local information has local effects.}

\noindent Suppose that the information to be processed refers only to a
subdomain $D$ of $X$ and nothing is said about values of $x$ outside $D$. We
design the inference method so that the probability for any $x$ conditional
on its being outside $D$, $p(x|x\notin D)$ is not updated. We emphasize: the
point is not that we make the unwarranted assumption that keeping $%
p(x|x\notin D)$ fixed will lead to correct inferences; it may not. The point
is, rather, that in the absence of any supporting evidence there is no
reason to change our minds.

The consequence of axiom 1 is that non-overlapping domains of $x$ contribute
additively to the entropy, 
\begin{equation}
S[p]=\int dx\,F\left( p(x),x\right) \ ,  \label{axiom1}
\end{equation}%
where $F$ is some unknown function.

\textbf{Axiom 2: Coordinate invariance.} \emph{The system of coordinates
carries no information. }

\noindent The points $x$ can be labeled using any of a variety of coordinate
systems. In certain situations we might have explicit reasons to believe
that a particular choice of coordinates should be preferred over others.
This information might have been given to us in a variety of ways, but
unless the evidence was, in fact, given, we should not assume it: the
ranking of probability distributions should not depend on the coordinates
used.

It may be useful to recall some facts about coordinate transformations.
Consider a change from old coordinates $x$ to new coordinates $x^{\prime }$
such that $x=\Gamma (x^{\prime })$. The new volume element $dx^{\prime }$
includes the corresponding Jacobian, 
\begin{equation}
dx=\gamma (x^{\prime })dx^{\prime }\quad \text{where}\quad \gamma (x^{\prime
})=\left\vert \frac{\partial x}{\partial x^{\prime }}\right\vert .
\label{coord jacobian}
\end{equation}%
Let $m(x)$ be any density; in the new coordinates it transforms so that $%
m(x)dx=m^{\prime }(x^{\prime })dx^{\prime }$. This is true, in particular,
for the probability density $p(x)$, therefore 
\begin{equation}
m^{\prime }(x^{\prime })=m\left( \Gamma (x^{\prime })\right) \gamma
(x^{\prime })\quad \text{and}\quad p^{\prime }(x^{\prime })=p\left( \Gamma
(x^{\prime })\right) \gamma (x^{\prime }).  \label{coord trans dens}
\end{equation}%
The coordinate transformation gives

\begin{equation}
S[p]=\int dx\,F\left( p(x),x\right) =\int \gamma (x^{\prime })dx^{\prime
}\,F\left( \frac{p^{\prime }(x^{\prime })}{\gamma (x^{\prime })},\Gamma
(x^{\prime })\right) ,
\end{equation}%
which is a mere change of variables. The identity above is valid always, for
all $\Gamma $ and for all $F$; it imposes no constraint on $S[p]$. The
constraint arises from realizing that we could equally well have ranked
distributions according to $S[p^{\prime }]=\int dx^{\prime }\,F\left(
p^{\prime }(x^{\prime }),x^{\prime }\right) $ and that this should have no
effect on our conclusions. This is the nontrivial constraint. It is not that
we can change variables, we can always do that; but rather that the two
rankings, the one according to $S[p]$ and the other according to $%
S[p^{\prime }]$ must coincide. This requirement is satisfied if, for
example, $S[p]$ and $S[p^{\prime }]$ turn out to be numerically equal, but
this is not necessary.

The consequence of axiom 2 is that $S[p]$ can be written in terms of
coordinate invariants such as $dx\,m(x)$ and $p(x)/m(x),$%
\begin{equation}
S[p]=\int dx\,m(x)\Phi \left( \frac{p(x)}{m(x)}\right) ~.  \label{axiom2}
\end{equation}

The density $m(x)$ and the function $\Phi $ are, at this point, still
undetermined. On the other hand the purpose for introducing $S$ in the first
place was to update from a prior $q(x)$ to a posterior $p(x)$. We expect the
entropy to be a functional $S[p|q]$ and not just $S[p]$. The question `Where
is the prior?' is answered by invoking the locality axiom once again. The
situation when no new information is available is a special case of the
situation when information is given about states in a domain $D$. When we
allow the domain $D$ to shrink to $\varnothing $ the requirement that the
conditional probabilities $p(x|x\notin D)=p(x|x\in X)=p(x)$ should not be
updated translates into

\textbf{Axiom 1 (special case): }\emph{When there is no new information
there is no reason to change one's mind. }

\noindent When there are no constraints the selected posterior distribution
should coincide with the prior distribution. The consequence of this second
use of locality is that the arbitrariness in the density $m(x)$ is removed:
up to normalization $m(x)$ is the prior distribution.

\textbf{Axiom 3:\ Subsystem independence}. \emph{When a system is composed
of subsystems that are believed to be independent it should not matter
whether the inference procedure treats them separately or jointly.}

\noindent Consider a system composed of two subsystems, $x=(x_{1},x_{2})\in
X=X_{1}\times X_{2}$. Assume that all prior evidence led us to believe the
systems were independent. This belief is reflected in the prior
distribution: if the subsystem priors $m_{1}(x_{1})$ and $m_{2}(x_{2})$,
then the prior for the whole system is $m_{1}(x_{1})m_{2}(x_{2})$. Further
suppose that new information is acquired such that $m_{1}(x_{1})$ is updated
to $p_{1}(x_{1})$ and $m_{2}(x_{2})$ is updated to $p_{2}(x_{2})$. Nothing
in this new information requires us to revise our previous assessment of
independence, therefore there is no need to change our minds, and the prior
for the whole system $m_{1}(x_{1})m_{2}(x_{2})$ should be updated to $%
p_{1}(x_{1})p_{2}(x_{2})$.

We emphasize that the point is not that when we have no evidence for
correlations we draw the firm conclusion that the systems must necessarily
be independent. They could indeed have turned out to be correlated and then
our inferences would be wrong. Induction involves some risk. The point is
rather that if we originally \emph{believe} the subsystems to be independent
and if the new evidence is silent on the matter of correlations, then there
is no reason to change our minds. Indeed, to the extent that we place any
value at all on whatever old evidence led us to believe the subsystems were
independent, then we ought not to change our minds. As before, a feature of
the probability distribution -- in this case, independence -- will not be
updated unless the evidence requires it.

The consequence of axiom 3 is to fix the function $\Phi $. The final
conclusion is that probability distributions $p(x)$ should be ranked
relative to the prior $m(x)$ according to their (relative) entropy, 
\begin{equation}
S[p|m]=-\int dx\,p(x)\log \frac{p(x)}{m(x)}.  \label{S[p]}
\end{equation}%
The derivation has singled out a unique $S[p|m]$ to be used in inductive
inference. Other expressions, may be useful for other purposes, but they do
not constitute an induction from the simple cases described in the axioms.
Of course, as emphasized above, induction is risky and failure is possible.
The most common cause of failure is that the constraints that are relevant
have not been properly identified. But it could very well happen that other
equally compelling axioms, leading to a different entropy, should have been
used as the basis for induction. An example is given below (Section 6).

\section{The proofs}

In this section we establish the consequences of the three axioms leading to
the final result eq.(\ref{S[p]}). The details of the proofs are important
not just because they lead to our final conclusions, but also because the
translation of the verbal statement of the axioms into precise mathematical
form is a crucial part of unambiguously specifying what the axioms actually
say. Uffink, for example, has shown \cite{Uffink95} how the same axioms of
Shore and Johnson \cite{ShoreJohnson80} which led them to the usual relative
entropy can be implemented mathematically in such a way that they lead not
to the usual relative entropy but rather to the Renyi entropies.

\subsection{Locality}

Here we prove that axiom 1 leads to the expression eq.(\ref{axiom1}) for $%
S[p]$. The requirement that probabilities be normalized is an annoying
technical complication. This problem can be conveniently overcome by
allowing the functional $S[p]$ to be defined for all functions with $%
p(x)\geq 0$ and \ to treat normalization as one among so many other
constraints that one might wish to impose.

To simplify the proof we consider the case of a discrete variable, $p_{i}$
with $i=1,\ldots ,n$, so that $S[p]=S(p_{1},\ldots ,p_{n})$. The
generalization to a continuum is straightforward.

Suppose the space of states $X$ is partitioned into two non-overlapping
domains $D$ and $D^{\prime }$ with $D\cup D^{\prime }=X$, and that the
information to be processed is in the form of separate constraints in each
domain, 
\begin{equation}
\sum_{i\in D}a_{i}p_{i}=A\quad \text{and \quad }\sum_{i\in D^{\prime
}}a_{i}^{\prime }p_{i}=A^{\prime }\text{ .}  \label{loc constr 1}
\end{equation}%
Axiom 1 states that the constraint on $D^{\prime }$ does not have an
influence on the conditional probabilities $p_{i|D}$. It may however
influence the $p_{i}$s within $D$ through an overall multiplicative factor.
To deal with this complication consider then a special case where the
overall probabilities of $D$ and $D^{\prime }$ are constrained too, 
\begin{equation}
\sum_{i\in D}p_{i}=P_{D}\quad \text{and \quad }\sum_{i\in D^{\prime
}}p_{i}=P_{D^{\prime }}\text{ ,}  \label{loc constr 2}
\end{equation}%
with $P_{D}+P_{D^{\prime }}=1$. Under these special circumstances
constraints on $D^{\prime }$ may not influence $p_{i}$s within $D$, and vice
versa.

The obtain the posterior maximize $S[p]$ subject to these four constraints, 
\begin{eqnarray*}
0 &=&\left[ \delta S-\lambda \left( \sum_{i\in D}p_{i}-P_{D}\right) +\mu
\left( \sum_{i\in D}a_{i}p_{i}-A\right) \right. \\
&&-\left. \lambda ^{\prime }\left( \sum_{i\in D^{\prime }}p_{i}-P_{D^{\prime
}}\right) +\mu ^{\prime }\left( \sum_{i\in D^{\prime }}a_{i}^{\prime
}p_{i}-A^{\prime }\right) \right] ~,
\end{eqnarray*}%
leading to 
\begin{eqnarray}
\frac{\partial S}{\partial p_{i}} &=&\lambda +\mu a_{i}\text{\quad for\quad }%
i\in D~,  \label{loc var} \\
\frac{\partial S}{\partial p_{i}} &=&\lambda ^{\prime }+\mu ^{\prime
}a_{i}^{\prime }\text{\quad for\quad }i\in D^{\prime }~.
\label{loc var prime}
\end{eqnarray}%
Eqs.(\ref{loc constr 1}-\ref{loc var prime}) are $n+4$ equations we must
solve for the $p_{i}$s and the four Lagrange multipliers.

Since $S=S(p_{1},\ldots ,p_{n})$ its derivative $\partial S/\partial
p_{i}=f_{i}(p_{1},\ldots ,p_{n})$ could in principle also depend on all $n$
variables. But this violates the locality axiom because any arbitrary change
in $a_{i}^{\prime }$ within $D^{\prime }$ would influence probabilities
outside $D^{\prime }$. The only way that probabilities within $D$ can be
shielded from arbitrary changes in the constraints pertaining to $D^{\prime
} $ is that the function $f_{i}(p_{1},\ldots ,p_{n})$ with $i\in D$ be
independent of all $p_{j}$'s with $j\in D^{\prime }$.

With this restriction on the function $f_{i}$ the two systems of equations
referring to $D$ and to $D^{\prime }$ become totally decoupled and locality
is preserved. But the decoupling must hold not just for one particular
partition of $X$ into domains $D$ and $D^{\prime }$, it must hold for all
conceivable partitions.\ Therefore the locality axiom requires $S[p]$ to be
such that its derivative $f_{i}$ depends only on the single variable $p_{i}$%
, 
\begin{equation}
\frac{\partial S}{\partial p_{i}}=f_{i}(p_{i})\quad \text{or}\quad \frac{%
\partial ^{2}S}{\partial p_{i}\partial p_{j}}=0\quad \text{for}\quad i\neq j.
\end{equation}%
Integrating, one obtains 
\begin{equation}
S[p]=\sum_{i}F_{i}(p_{i})+\func{constant}\text{.}
\end{equation}%
for some undetermined functions $F_{i}$. The corresponding expression for a
continuous variable $x$ is obtained replacing $i$ by $x$, and the sum over $%
i $ by an integral over $x$ leading to eq.(\ref{axiom1}).

\subsection{Coordinate invariance}

Next we prove eq.(\ref{axiom2}) It is convenient to introduce a function $%
m(x)$ which transforms as a density and rewrite the expression (\ref{axiom1}%
) for the entropy in the form 
\begin{equation}
S[p]=\int dx\,m(x)\frac{1}{m(x)}F\left( \frac{p(x)}{m(x)}m(x),x\right) =\int
dx\,m(x)\Phi \left( \frac{p(x)}{m(x)},m(x),x\right) ,
\end{equation}%
where the function $\Phi $ is defined by 
\begin{equation}
\Phi (\alpha ,m,x)\overset{\limfunc{def}}{=}\frac{1}{m}F(\alpha m,x).
\end{equation}

Next, we consider a special situation where the new information are
constraints which do not favor one coordinate system over another. For
example consider the constraint 
\begin{equation}
\int dx\,p(x)a(x)=A
\end{equation}%
where $a(x)$ is a scalar, \emph{i.e.}, invariant under coordinate changes, 
\begin{equation}
a(x)\rightarrow a^{\prime }(x^{\prime })=a(x).  \label{ci-a}
\end{equation}%
The usual normalization condition $\int dx\,p(x)=1$ is a simple example of a
scalar constraint.

Maximizing $S[p]$ subject to the constraint, 
\begin{equation}
\delta \left[ S[p]+\lambda \left( \int dx\,p(x)a(x)-A\right) \right] =0,
\end{equation}%
gives 
\begin{equation}
\dot{\Phi}\left( \frac{p(x)}{m(x)},m(x),x\right) =\lambda a(x)~,
\label{ci-b}
\end{equation}%
where 
\begin{equation}
\dot{\Phi}\left( \alpha ,m,x\right) \overset{\limfunc{def}}{=}\frac{\partial
\Phi \left( \alpha ,m,x\right) }{\partial \alpha }
\end{equation}%
is just the derivative with respect to the first argument. But we could have
started using the primed coordinates, 
\begin{equation}
\dot{\Phi}\left( \frac{p^{\prime }(x^{\prime })}{m^{\prime }(x^{\prime })}%
,m^{\prime }(x^{\prime }),x^{\prime }\right) =\lambda ^{\prime }a^{\prime
}(x^{\prime }),
\end{equation}%
or equivalently, using eqs.(\ref{coord trans dens}) and (\ref{ci-a}), 
\begin{equation}
\dot{\Phi}\left( \frac{p(x)}{m(x)},m(x)\gamma (x^{\prime }),x^{\prime
}\right) =\lambda ^{\prime }a(x).  \label{ci-c}
\end{equation}%
Dividing (\ref{ci-c}) by (\ref{ci-b}) we get 
\begin{equation}
\frac{\dot{\Phi}\left( \alpha ,m\gamma ,x^{\prime }\right) }{\dot{\Phi}%
\left( \alpha ,m,x\right) }=\frac{\lambda ^{\prime }}{\lambda }.
\label{ci-d}
\end{equation}%
This identity should hold for any transformation $x=\Gamma (x^{\prime })$.
On the right hand side the multipliers $\lambda $ and $\lambda ^{\prime }$
are just constants; the ratio $\lambda ^{\prime }/\lambda $ might depend on
the transformation $\Gamma $ but it does not depend on $x$. Consider the
special case of a transformation $\Gamma $ that has unit determinant
everywhere, $\gamma =1$, and differs from the identity transformation only
within some arbitrary region $D$. Since for $x$ outside this region $D$ we
have $x=x^{\prime }$, the left hand side of eq.(\ref{ci-d}) equals 1. Thus,
for this particular $\Gamma $ the ratio is $\lambda ^{\prime }/\lambda =1$;
but $\lambda ^{\prime }/\lambda =\func{constant}$, so $\lambda ^{\prime
}/\lambda =1$ holds within $D$ as well. Therefore, for $x$ within $D$, 
\begin{equation}
\dot{\Phi}\left( \alpha ,m,x^{\prime }\right) =\dot{\Phi}\left( \alpha
,m,x\right) .
\end{equation}%
Since the choice of $D$ is arbitrary we conclude is that the function $\dot{%
\Phi}$ cannot depend on its third argument, $\dot{\Phi}=\dot{\Phi}\left(
\alpha ,m\right) $.

Having eliminated the third argument, let us go back to eq.(\ref{ci-d}), 
\begin{equation}
\frac{\dot{\Phi}\left( \alpha ,m\gamma \right) }{\dot{\Phi}\left( \alpha
,m\right) }=\frac{\lambda ^{\prime }}{\lambda }\ ,
\end{equation}%
and consider a different transformation $\Gamma $, one with unit determinant 
$\gamma =1$ outside the region $D$. Therefore the constant ratio $\lambda
^{\prime }/\lambda $ is again equal to $1$, so that 
\begin{equation}
\dot{\Phi}\left( \alpha ,m\gamma \right) =\dot{\Phi}\left( \alpha ,m\right) .
\end{equation}%
But within $D$ the transformation $\Gamma $ is quite arbitrary, it could
have any arbitrary Jacobian $\gamma \neq 1$. Therefore the function $\dot{%
\Phi}$ cannot depend on its second argument either, $\dot{\Phi}=\dot{\Phi}%
(\alpha )$. Integrating with respect to $\alpha $ gives $\Phi =\Phi (\alpha
)+\func{constant}$. The additive constant has no effect on the maximization
and can be dropped. This completes the proof of eq.(\ref{axiom2}).

\subsection{The prior}

The locality axiom implies that when there are no constraints the selected
posterior distribution should coincide with the prior distribution. This
provides us with an interpretation of the measure $m(x)$ that had been so
artificially introduced. The argument is simple: maximize $S[p]$ in (\ref%
{axiom2}) subject to the single requirement of normalization, 
\begin{equation}
\delta \left[ S[p]+\lambda \left( \int dx\,p(x)-1\right) \right] =0,
\end{equation}%
to get 
\begin{equation}
\dot{\Phi}\left( \frac{p(x)}{m(x)}\right) =\lambda .  \label{sc-a}
\end{equation}%
Since $\lambda $ is a constant, the left hand side must be independent of $x$%
. This could, for example, be accomplished if the function $\dot{\Phi}%
(\alpha )$ were itself a constant, independent of its argument $\alpha $.
But this gives $\Phi (\alpha )=c_{1}\alpha +c_{2}$, where $c_{1}$ and $c_{2}$
are constants, and leads to the unacceptable form $S[p]\propto \int dx\,p(x)$%
. If the dependence on $x$ cannot be eliminated by an appropriate choice of $%
\dot{\Phi}$, we must secure it by a choice of $m(x)$. Eq.(\ref{sc-a}) is an
equation for $p(x)$; the obvious solution is $p(x)\propto m(x)$. But in the
absence of new information the selected posterior distribution must reflect
our prior beliefs, therefore $m(x)$ must, except for an overall
normalization, be chosen to coincide with the prior distribution.

\subsection{Independent subsystems}

If $x=(x_{1},x_{2})\in X=X_{1}\times X_{2}$, and the subsystem priors $%
m_{1}(x_{1})$ and $m_{2}(x_{2})$ are independently updated to $p_{1}(x_{1})$
and $p_{2}(x_{2})$ respectively, then the prior for the whole system $%
m_{1}(x_{1})m_{2}(x_{2})$ should be updated to $p_{1}(x_{1})p_{2}(x_{2})$.

We need only consider a special case where the posterior distributions for
the individual systems, $p_{1}(x_{1})$ and $p_{2}(x_{2})$, happen to be
known. When the systems are treated separately this is the trivial case of
extremely constraining information: for system $1$ we want to maximize $%
S_{1}[p]$ subject to the constraint that $p(x_{1})$ is $p_{1}(x_{1})$, the
result being, naturally, $p(x_{1})=p_{1}(x_{1})$. A similar result holds for
system $2$.

When the systems are treated jointly, however, the inference is not nearly
as trivial. We want to maximize the entropy of the joint system, 
\begin{equation}
S[p]=\int dx_{1}dx_{2}\,m(x_{1},x_{2})\Phi \left( \frac{p(x_{1},x_{2})}{%
m(x_{1},x_{2})}\right) ,
\end{equation}%
where the joint prior $m(x_{1},x_{2})$ is a product, $%
m_{1}(x_{1})m_{2}(x_{2})$, and the constraints on the joint distribution $%
p(x_{1},x_{2})$ are 
\begin{equation}
\int dx_{2}\,p(x_{1},x_{2})=p_{1}(x_{1})\qquad \text{and\qquad }\int
dx_{1}\,p(x_{1},x_{2})=p_{2}(x_{2}).
\end{equation}%
Notice that here we have not written just two constraints. We actually have
one constraint for each value of $x_{1}$ and of $x_{2}$; this is an infinity
of constraints, each of which must be multiplied by its own Lagrange
multiplier, $\lambda _{1}(x_{1})$ or $\lambda _{2}(x_{2})$. Then, 
\begin{equation}
\delta \left[ S[p]-\int dx_{1}\lambda _{1}(x_{1})\left( \int
dx_{2}\,p(x_{1},x_{2})-p_{1}(x_{1})\right) -\{1\leftrightarrow 2\}\right] =0,
\end{equation}%
where $\{1\leftrightarrow 2\}$ indicates a third term, similar to the
second, with $1$ and $2$ interchanged. The independent variations $\delta
p(x_{1},x_{2})$ yield 
\begin{equation}
\Phi ^{\prime }\left( \frac{p(x_{1},x_{2})}{m_{1}(x_{1})m_{2}(x_{2})}\right)
=\lambda _{1}(x_{1})+\lambda _{2}(x_{2}).
\end{equation}%
(The prime indicates a derivative with respect to the argument.) But we know
that the selected posterior should be the product $%
p(x_{1},x_{2})=p_{1}(x_{1})p_{2}(x_{2})$. Then, 
\begin{equation}
\Phi ^{\prime }\left( y\right) =\lambda _{1}(x_{1})+\lambda
_{2}(x_{2}),\quad \text{where\quad }y=\frac{p_{1}(x_{1})p_{2}(x_{2})}{%
m_{1}(x_{1})m_{2}(x_{2})}.  \label{is-b}
\end{equation}%
Differentiating with respect to $x_{1}$ and to $x_{2}$, yields 
\begin{equation}
y\Phi ^{\prime \prime \prime }(y)+\Phi ^{\prime \prime }(y)=0~,
\end{equation}%
which can easily be integrated three times to give 
\begin{equation}
\Phi (y)=ay\log y+by+c.
\end{equation}%
The additive constant $c$ may be dropped: its contribution to the entropy
would appear in a term that does not depend on the probabilities and would
have no effect on the ranking scheme. At this point the entropy takes the
form 
\begin{equation}
S[p]=\int dx\,\left( ap(x)\log \frac{p(x)}{m(x)}+bp(x)\right) .
\end{equation}%
This $S[p]$ will be maximized subject to constraints which always include
normalization. Since this is implemented by adding a term $\lambda \int
dx\,p(x)$, the $b$ constant can always be absorbed into the undetermined
multiplier $\lambda .$ Thus, the term $bp(x)$ has no effect on the selected
distribution and can be dropped.

Finally, $a$ is just an overall multiplicative constant, it also does not
affect the overall ranking except in the trivial sense that inverting the
sign of $a$ will transform the maximization problem to a minimization
problem or vice versa. We can therefore set $a=-1$ so that maximum $S$
corresponds to maximum preference. The opposite choice $a=1$ leads to what
is usually called the cross-entropy or the Kullback number.

\section{To what extent are non-ME distributions ruled out?}

Suppose we have maximized the entropy (\ref{S[p]}) subject to certain
constraints and obtained a probability distribution $p_{0}(x)$. The question
we now address concerns the extent to which $p_{0}(x)$ should be preferred
over other distributions with lower entropy. Consider a family of
distributions $p(x|\theta )$, labelled by a finite number $n_{{}\theta }$ of
parameters $\theta ^{i}$ ($i=1,\ldots ,n_{{}\theta }$). We assume that the $%
p(x|\theta )$ satisfy the same constraints and include $p_{0}(x)=p(x|\theta
=0)$.

The question about the extent to which $p(x|\theta =0)$ is preferred over $%
p(x|\theta \neq 0)$ is a question about the probability of $\theta $, $\pi
(\theta )$. The original problem which led us to invoke ME method was to
assign a probability to $x$; our new problem is to assign probabilities to $%
x $ and $\theta $.\ We are concerned not just with $p(x)$ but rather with $%
p(x,\theta )$; the universe of discourse has been expanded from $X$ to $%
X\times \Theta $ where $\Theta $ is the space of parameters $\theta $. The
joint distribution $p(x,\theta )$ will also be determined using the ME
method. To proceed we must address two questions: What is the prior
distribution, what do we know about $x$ and $\theta $ before we learnt about
the constraints? And second, what are the constraints on $p(x,\theta )$?

This first question is the more subtle one: when we know nothing about the $%
\theta $s we know neither their physical meaning nor whether there is any
relation to the $x$. A prior that reflects this lack of correlations is a
product, $m(x,\theta )=m(x)\mu (\theta )$. We will assume that the prior
over $x$ is known (it is the prior we had used to update from $m(x)$ to $%
p_{0}(x)$), but $\mu (\theta )$\ is unknown. Suppose next that we are told
that the $\theta $s are just parameters labeling some distributions $%
p(x|\theta )$. We do not yet know the functional form of $p(x|\theta )$, but
if the $\theta $s derive their meaning solely from the $p(x|\theta )$ then
for each choice of $p(x|\theta )$ there is a natural distance in the space $%
\Theta $: it is given by the Fisher-Rao metric $d\ell ^{2}=g_{ij}d\theta
^{i}d\theta ^{j}$, \cite{Amari85} 
\begin{equation}
g_{ij}=\int dx\,p(x|\theta )\frac{\partial \log p(x|\theta )}{\partial
\theta ^{i}}\frac{\partial \log p(x|\theta )}{\partial \theta ^{j}}.
\label{Fisher metric}
\end{equation}%
Accordingly we choose $\mu (\theta )=g^{1/2}(\theta )$, where $g(\theta )$
is the determinant of $g_{ij}$.

To each different choice of the functional form of $p(x|\theta )$ there
corresponds a different subspace of the space of joint distributions defined
by distributions of the form $p(x,\theta )=\pi (\theta )p(x|\theta )$. The
crucial constraint specifies which particular functional form for $%
p(x|\theta )$ we have in mind; this provides meaning to the $\theta $s and
fixes the prior and the relevant subspace. Notice that this constraint is
not in the usual form of an expectation value.

The preferred distribution $p(x,\theta )$ is chosen by varying $\pi (\theta
) $ to maximize 
\begin{equation}
\sigma \lbrack \pi ]=-\int dx\,d\theta \,\pi (\theta )p(x|\theta )\,\log 
\frac{\pi (\theta )p(x|\theta )}{g^{1/2}(\theta )m(x)}=\mathcal{S}[\pi
]+\int d\theta \,\pi (\theta )S(\theta ),  \label{sigma[p]}
\end{equation}%
where 
\begin{equation}
\mathcal{S}[\pi ]=-\int \,d\theta \,\pi (\theta )\log \frac{\pi (\theta )}{%
g^{1/2}(\theta )}\quad \text{and}\quad S(\theta )=-\int \,dx\,p(x|\theta
)\log \frac{p(x|\theta )}{m(x)}.  \label{Stheta}
\end{equation}%
The notation shows that $\sigma \lbrack \pi ]$ and $\mathcal{S}[\pi ]$ are
functionals of $\pi (\theta )$ while $S(\theta )$ is a function of $\theta $%
. Maximizing (\ref{sigma[p]}) with respect to variations $\delta \pi (\theta
)$ such that $\int d\theta \,\pi (\theta )=1$, yields 
\begin{equation}
0=\int \,d\theta \left( -\log \frac{\pi (\theta )}{g^{1/2}(\theta )}%
+S(\theta )+\log \zeta \right) \,\delta \pi (\theta )\,,
\end{equation}%
where the required Lagrange multiplier has been written as $1-\log \zeta $.
Therefore the probability that the value of $\theta $ should lie within the
small volume $g^{1/2}(\theta )d\theta $ is 
\begin{equation}
\pi (\theta )d\theta =\frac{1}{\zeta }\,\,e^{S(\theta )}g^{1/2}(\theta
)d\theta \quad \text{with\quad }\zeta =\int d\theta \,g^{1/2}(\theta
)\,e^{S(\theta )}.  \label{main}
\end{equation}%
Equation (\ref{main}) is the result we seek. It tells us that, as expected,
the preferred value of $\theta $ is that which maximizes the entropy $%
S(\theta )$ because this maximizes the scalar probability density $\exp
S(\theta )$. But it also tells us the degree to which values of $\theta $
away from the maximum are ruled out. For macroscopic systems the preference
for the ME distribution can be overwhelming. Eq.(\ref{main}) agrees with the
Einstein thermodynamic fluctuation theory and extends it beyond the regime
of small fluctuations \cite{Caticha00}. Note also that the density $\exp
S(\theta )$ is a scalar function and the presence of the Jacobian factor $%
g^{1/2}(\theta )$ makes Eq.(\ref{main}) manifestly invariant under changes
of the coordinates $\theta ^{i}$ in the space $\Theta $.

To conclude this section we remark that there is a certain analogy in the
relation between the MaxEnt and ME methods and the relation between the
maximum likelihood and Bayes' theorem methods. Maximizing the likelihood
function $L(\theta |x)\overset{\limfunc{def}}{=}p(x|\theta )$ selects a
single preferred $\theta $. But $L(\theta |x)$ is not a probability
distribution for $\theta $ and the maximum likelyhood method does not,
without further elaborations and unlike the more general Bayesian approach,
address the question of the extent to which other values of $\theta $ are
ruled out.

\section{Random remarks}

\subsection{\noindent Choosing the prior}

Choosing the prior density $m(x)$ can be tricky. When there is no
information leading us to prefer one microstate of a physical system over
another we might as well assign equal prior probability to each state. Thus
it is reasonable to identify $m(x)$ with the density of states and the
invariant $m(x)dx$ is the number of microstates in $dx$. This is the basis
for statistical mechanics -- the postulate of equal a priori probabilities.
Other examples of relevance to physics arise when there is no reason to
prefer one region of the space $X$ over another. Then we should assign the
same prior probability to regions of the same volume,\ and we can choose $%
\int_{R}dx\,m(x)$ to be the volume of a region $R$ in the space $X$.

\emph{All entropies are relative entropies}. In the case of a discrete
variable, if we assign equal a priori probabilities, $m_{i}=1$, the entropy
is 
\begin{equation}
S[p]=-\sum_{i}\,p_{i}\log p_{i}\,,  \label{S[p]1}
\end{equation}%
the entropy function discovered by Boltzmann and by Shannon. The notation $%
S[p]$ has a serious drawback: it misleads one into thinking that $S$ depends
on $p(x)$ only. In particular, we emphasize that whenever the expression (%
\ref{S[p]1}) is used, the prior measure $m_{i}=1$ has been implicitly
assumed. In Shannon's axioms, for example, this choice is implicitly made in
his first axiom, when he states that the entropy is a function of the
probabilities $S=S(p_{1}...p_{n})$ and nothing else, and also in his second
axiom when the uniform distribution $p_{i}=1/n$ is singled out for special
treatment.

The absence of an explicit reference to a prior $m_{i}$ in (\ref{S[p]1}) may
erroneously suggest that prior distributions have been rendered unnecessary
and can be eliminated. It suggests that it is possible to transform
information (\emph{i.e.}, constraints) directly into posterior distributions
in a totally objective and unique way. If this were true the old
controversy, of whether probabilities are subjective or objective, would
have been resolved -- probabilities would ultimately be totally objective.
But the prior $m_{i}=1$ is implicit in eq.(\ref{S[p]1}); the postulate of
equal a priori probabilities or Laplace's \textquotedblleft Principle of
Insufficient Reason\textquotedblright\ still plays a major, though perhaps
hidden, role. Any claims that probabilities assigned using maximum entropy
will yield absolutely objective results are unfounded; not all subjectivity
has been eliminated. Just as with Bayes' theorem, what is objective here is
the manner in which information is processed to update from a prior to a
posterior, and not the prior probability assignments themselves.

What if $m(x)=0$ for some $x$? $S[p|m]$ can be infinitely negative when $%
m(x) $ vanishes within some region $D$. In other words, the ME method
confers an overwhelming preference on those distributions $p(x)$ that vanish
whenever $m(x)$ does. But this is not a problem. A similar \textquotedblleft
problem\textquotedblright\ also arises in the context of Bayes' theorem
where a vanishing prior represents a tremendously serious prejudice because
no amount of data to the contrary would allow us to revise it. The solution
in both cases is to recognize that unless we are absolutely certain that $x$
could not possibly lie within $D$ then we should not have assigned $m(x)=0$
in the first place. Assigning a very low but non-zero prior represents a
safer and less prejudiced representation of one's beliefs both in the
context of Bayesian and of ME inference.

\subsection{Entropy as a measure of information}

The notion of information is a vague one. Any attempt to find its measure
will always be open to the objection that it is not clear what is being
measured. On the other hand there is absolutely no ambiguity involved in the
prescription of how entropy as preference is used -- even if one does not
know precisely what is being preferred. It is a prescription motivated by
one's desire not to change one's mind unless compelled by concrete evidence.
It appears that rather than allowing the vagueness of the notion of amount
of information to contaminate the notion of entropy one should proceed in
the other direction and allow the unambiguous notion of entropy to confer
precision on the notion of amount of information.

Thus the amount of information missing in a discrete distribution $p_{i}$
should be defined as a relative entropy $S[p|m]$. Since $S[p|m]$ is
maximized for $p_{i}\propto m_{i}$, the special distribution $m_{i}$ should
be selected to agree with whatever prior notions we have about which
distribution contains the least information. A reasonable candidate,
suggested by the Principle of Insufficient Reason, is the uniform
distribution, $m_{i}=m$. The constant $m$ is then determined by the
requirement that when we have complete knowledge there is no missing
information. Imposing $S[p|m]=0$ when $p_{i}=\delta _{ij}$ for some integer $%
j$, yields $m=1$. Thus Shannon's measure is recovered.

\subsection{Comments on other axiomatizations}

One feature that distinguishes the various axiomatizations is how they
justify maximizing a functional. In other words why \emph{maximum }entropy?
In the approach of Shore and Johnson this question receives no answer; it is
just one of the axioms. Csiszar provides a better answer. He derives the
`maximize a functional' rule from reasonable axioms of regularity and
locality \cite{Csiszar91}. In Skilling's and in the approach developed here
the rule is not derived, but it does no go unexplained either: it is imposed
by design, it is justified by the function that $S$ is supposed to perform,
to achieve a transitive ranking.

Both Shore and Johnson and Csiszar require, and it is not clear why, that
updating from a prior must lead to a unique posterior, and accordingly,
there is a restriction that the constraints define a convex set. In
Skilling's approach and in the one advocated here there is no requirement of
uniqueness, we are perfectly willing to entertain situations where the
available information points to several equally preferable distributions.

There is an important difference between the axiomatic approach presented by
Csiszar and the present one. Since our ME method is a method for induction
we are justified in applying the method as if it were of universal
applicability. As with all inductive procedures, in any particular instance
of induction can turn out to be wrong -- because, for example, not all
relevant information has been taken into account -- but this does not change
the fact that ME is still the unique inductive inference method that
generalizes from the special cases chosen as axioms. Csiszar's version of
the MaxEnt method is not designed to generalize beyond the axioms. His
method was developed for linear constraints and therefore he does not feel
justified in carrying out his \emph{deductions} beyond the cases of linear
constraints. In our case, the form of $S[p|q]$ from the axioms was a matter
of \emph{deduction} but the application to non-linear constraints is
precisely the kind of \emph{induction} we want to carry out.

\subsection{On constraints}

First of all, one should not confuse questions about how information should
be processed from questions about how the information is obtained. This
applies both to the case of processing data using Bayes' theorem and of
processing information in the form of constraints or testable information
using ME. Bayes' theorem solves the problem of how data is used to update
from a prior to a posterior distribution; it does not address all those
interesting issues concerning the actual collection of data -- the whole of
experimental science. Similarly, the ME method is designed to process
information in the form of a specification of the family of allowed
posteriors. Where and how that information is obtained is not a problem
addressed by the ME method.

Having made that distinction, we can still ask how the information to be
processed using ME is actually obtained. One point to be made is that
empirical data, even sample averages, do not refer to probabilities, and
therefore do not provide constraints on probability distributions. Confusing
expected values with sample averages leads to inconsistencies, particularly
for small samples \cite{Uffink96}. Data on sample averages requires Bayes'
theorem, information about expected values requires ME. Of course, for very
large samples inconsistencies disappear, and both the Bayesian and the ME
approaches agree.

Once we accept that constraints will refer to the expected values of certain
variables, how do we decide their numerical magnitudes? And, for that
matter, which variables do we choose? Indeed, what constraints should we
choose?

When justifying the use of the ME method to obtain, say, the canonical
Boltzmann factors ($P_{q}\propto e^{-\beta E_{q}}$) it has been common to
say something like \textquotedblleft we seek the minimally biased (\emph{i.e.%
} maximum entropy) distribution that codifies the information we have (the
expected energy) and nothing else\textquotedblright . Many authors find this
justification objectionable. Indeed, they might argue, for example, that the
spectrum of black body radiation is what it is independently of whatever
information happens to be available to us. We prefer to phrase the objection
differently: in most realistic situations the expected value of the energy
is not a quantity we happen to know. Nevertheless, it is still true that
maximizing entropy subject to a constraint on this (unknown) expected energy
leads to the right family of distributions. Therefore, the justification
behind imposing a constraint on the expected energy cannot be that this is a
quantity that happens to be known -- because of the brute fact that we never
know it -- but rather because the\emph{\ }expected energy is the quantity
that \emph{should }be\emph{\ }known. Even if unknown, we recognize it as the
crucial relevant information without which no successful predictions are
possible. Therefore we proceed as if this crucial information were available
and produce a formalism that contains the temperature as a free parameter.
The actual value of the temperature will have to be inferred from the
experiment itself either directly, using a thermometer, or indirectly by
Bayesian analysis from other empirical data.

To summarize: the constraints that should be imposed are those that codify
information that, even if unknown, is relevant and necessary for a
successful inference.

One last remark on constraints and their relation to priors: \noindent the
distribution $m(x)$ represents our prior beliefs, including information that
might have been taken into account in earlier applications of the ME method.
It is important to realize that later applications of ME for processing new
constraints need not in general preserve old constraints. The reason is that
when a new constraint is given one is implicitly admitting that all
probability distributions satisfying the new constraint are in principle
possible, and this will be interpreted as evidence which contradicts the old
constraints and requires their updating. The specification of the allowed
family of posteriors must be complete: this means that in addition to the
new constraints one should also impose those old constraints that are not
meant to be updated.

\section{Is entropy unique?}

The entropy is a tool for generalizing from special cases and its functional
form follows from the choice of the specific cases described in the axioms.
One could very well expect that a different choice of special cases would
lead to a different generalization. Is entropy unique? Is there a universal
theory of inductive inference for processing testable information? Here we
develop a theory of inference based on a different choice for the third
axiom which is the one that determines the function $\Phi $ in eq.(\ref%
{axiom2}).

\subsection{An alternative third axiom}

Consider a system composed of two subsystems, $x=(x_{1},x_{2})\in
X=X_{1}\times X_{2}$. Assume that prior evidence has led us to believe the
subsystems are not independent. This belief is reflected in a prior
distribution $m(x_{1},x_{2})$ which does not factor into a product of
independent priors for the subsystems. Beyond telling us how we believe the
subsystems are correlated, the prior $m(x_{1},x_{2})$ also tells us what we
believe about each of the two subsystems separately. These beliefs are
codified in the marginal distributions 
\begin{equation}
\int dx_{2}~m(x_{1},x_{2})=m_{1}(x_{1})\quad \text{and\quad }\int
dx_{1}~m(x_{1},x_{2})=m_{2}(x_{2})~.  \label{marginals}
\end{equation}%
Now suppose that new information tells us that the two subsystems are
actually independent. We want to select the posterior within the family of
independent distributions, $p_{1}(x_{1})p_{2}(x_{2})$. Thus we should update
our old information about correlations, but since the new evidence is silent
about those aspects of the distribution that refer to each of the subsystems
by themselves there is no reason to update the marginal distributions. In
this case the updated posterior should be $%
p(x_{1},x_{2})=m_{1}(x_{1})m_{2}(x_{2})$. This is written as an alternative
third axiom:

\textbf{Axiom 3-alt:\ Subsystem marginals}. \emph{When a system is composed
of subsystems and the information to be processed refers }only\emph{\ to the
correlations between them there is no need to update whatever beliefs we
might have about them individually. }

\noindent We emphasize yet again that the point is not that when we have no
evidence that requires updating the marginals we conclude they must
necessarily remain unchanged. It is just that a feature of the probability
distribution -- in this case, the marginals -- will not be updated unless
the evidence requires it.

The consequence of Axiom 3-alt is to fix the function $\Phi $. The final
conclusion is that probability distributions $p(x)$ should be ranked
relative to the prior $m(x)$ according to a new (relative) entropy, 
\begin{equation}
\tilde{S}[p|m]=\int dx\,m(x)\log \frac{p(x)}{m(x)}~,
\end{equation}%
which, incidentally, happens to be the \textquotedblleft
dual\textquotedblright\ of the relative entropy: $\tilde{S}[p|m]=S[m|p]$.

Before we proceed to the proof we remark on the significance of this result:
The two axioms 3 and 3-alt seem equally compelling in the sense that both
refuse to update features of the prior unless required by the evidence. But
the two axioms are incompatible with each other. Therefore, we are led to
conclude that there is no general theory of induction which simultaneously
applies to the special cases described in all four of the axioms.

\subsection{Proof}

The mathematical manipulations below follow closely those used by Skilling
to solve a related but decidedly different problem -- that of selecting a
model \cite{Skilling88}.

Maximize the joint entropy 
\begin{equation}
\tilde{S}[p]=\int dx_{1}dx_{2}\,m(x_{1},x_{2})\Phi \left( \frac{%
p(x_{1},x_{2})}{m(x_{1},x_{2})}\right) ,
\end{equation}%
subject to the constraint that $p(x_{1},x_{2})=p_{1}(x_{1})p_{2}(x_{2})$,
and that $p_{1}(x_{1})$ and $p_{2}(x_{2})$ are individually normalized.
Independent variations $\delta p_{1}(x_{1})$ and $\delta p_{2}(x_{2})$ lead
to 
\begin{equation}
\int dx_{2}\,\Phi ^{\prime }\left( \frac{p_{1}(x_{1})p_{2}(x_{2})}{%
m(x_{1},x_{2})}\right) p_{2}(x_{2})=\lambda _{1}\quad \text{and\quad }%
\{1\leftrightarrow 2\}~,
\end{equation}%
where the prime denotes derivative with respect to the argument and $%
\{1\leftrightarrow 2\}$ denotes a similar equation with the subscripts $1$
and $2$ interchanged. We know that the selected posterior should be $%
m_{1}(x_{1})m_{2}(x_{2})$, therefore we obtain the following equations for
the function $\Phi $, 
\begin{equation}
\int dx_{2}\,\Phi ^{\prime }\left( \frac{m_{1}m_{2}}{m}\right) p_{2}=\lambda
_{1}\quad \text{and\quad }\{1\leftrightarrow 2\}~,
\end{equation}%
where the arguments $x_{1}$ and $x_{2}$ have been omitted. But there are
many other priors $M=m+\delta m$ with exactly the same marginals which
should lead to the same inference. Changing the prior by $\delta m$, changes
the equation for $\Phi $ by 
\begin{equation}
\int dx_{2}\,\Phi ^{\prime \prime }\left( \frac{m_{1}m_{2}}{m}\right) \left( 
\frac{m_{1}m_{2}}{m}\right) ^{2}\delta m=-m_{1}\delta \lambda _{1}\quad 
\text{and\quad }\{1\leftrightarrow 2\}~.  \label{eq for Phi}
\end{equation}

The conditions for $\delta m$ to preserve the marginals are that%
\begin{equation}
\int dx_{1}\,\delta m(x_{1},x_{2})=0\quad \text{for all }x_{2}\text{, and}%
\quad \{1\leftrightarrow 2\}~.
\end{equation}%
Since the most general marginal-preserving\ $\delta m$ is given by
perturbations of the form 
\begin{eqnarray}
\delta m(x_{1},x_{2}) &=&\int da_{1}db_{1}da_{2}db_{2}~\varepsilon
(a_{1},b_{1},a_{2},b_{2})  \notag \\
&&\left[ \delta (x_{1}-a_{1})-\delta (x_{1}-b_{1})\right] \left[ \delta
(x_{2}-a_{2})-\delta (x_{2}-b_{2})\right] ~,
\end{eqnarray}%
we need only consider the special case%
\begin{equation}
\delta m(x_{1},x_{2})=\varepsilon \left[ \delta (x_{1}-a_{1})-\delta
(x_{1}-b_{1})\right] \left[ \delta (x_{2}-a_{2})-\delta (x_{2}-b_{2})\right]
~.  \label{delta m}
\end{equation}%
Substituting into eq.(\ref{eq for Phi}) gives 
\begin{equation}
\left[ \delta (x_{1}-a_{1})-\delta (x_{1}-b_{1})\right] \left. \Phi ^{\prime
\prime }\left( \frac{m_{1}m_{2}}{m}\right) \left( \frac{m_{1}m_{2}}{m}%
\right) ^{2}\right\vert _{b_{2}}^{a_{2}}=-m_{1}(x_{1})\delta \lambda _{1}~,
\end{equation}%
and\ $\{1\leftrightarrow 2\}$. In order for this equation to hold when the $%
\delta $ functions vanish ($x_{1}\neq a_{1}$ and $x_{1}\neq b_{1}$) we must
have $\delta \lambda _{1}=0$. In order to hold when the $\delta $ functions
do not vanish, for example when $x_{1}=a_{1}$, we must have 
\begin{equation}
\left. \Phi ^{\prime \prime }\left( \frac{m_{1}(a_{1})m_{2}(x_{2})}{%
m(a_{1},x_{2})}\right) \left( \frac{m_{1}(a_{1})m_{2}(x_{2})}{m(a_{1},x_{2})}%
\right) ^{2}\right\vert _{x_{2}=b_{2}}^{x_{2}=a_{2}}=0~.
\end{equation}%
The only way this equation can hold for arbitrary choices of $a_{1}$, $a_{2}$
and $b_{2}$ is that the function $\Phi $ be such that 
\begin{equation}
\Phi ^{\prime \prime }(y)y^{2}=A=\func{constant}
\end{equation}%
Integrating twice leads to 
\begin{equation}
\Phi ^{\prime }(y)=-\frac{A}{y}+B\quad \text{and}\quad \Phi (y)=-A\log
y+By+C~,
\end{equation}%
so that, 
\begin{equation}
\tilde{S}[p|m]=-A\int dx\,m(x)\log \frac{p(x)}{m(x)}+B\int dx\,p(x)+C\int
dx\,m(x)~.
\end{equation}%
The last term is a constant independent of $p(x)$, it has no influence on
the ranking of $p$s and can be dropped. The constant $B$ can always be
absorbed into the Lagrange multiplier for the normalization constraint; it
can be dropped too. Finally, we choose $A=-1$ so that maximum $\tilde{S}%
[p|m] $ corresponds to maximum preference. This concludes the proof.

\section{Conclusions and final remarks}

We have established that the ME is a full-fledged method for updating from
prior to posterior distributions. It is designed for the processing of
testable information. Furthermore, the method does not just determine a
single posterior but it allows one to quantify the extent to which other
distributions that also satisfy the constraints are ruled out.

An important feature of ME is that the entropy requires no interpretation;
it is merely a tool for updating. Its functional form, however, depends on
the specific choice of cases from which one intends to generalize. Thus,
there is no general theory of induction -- not for probability distributions
and much less for positive functions. Having found a second apparently
legitimate entropy, the door is open to the possibility that there may be
others. Indeed, it appears that legislating that one should not update any
features of the prior except when forced by the evidence is too restricting:
we must accept the fact that not all features of the prior can be preserved,
that some features take precedence over others. Different choices of axioms
correspond to different choices of which features are preferred.

It is an empirical fact that selecting independence as a feature to be
preferred leads to correct inferences in an enormously wide variety of cases
including the whole of range of systems successfully described in
statistical physics and physical chemistry -- this includes all sorts of
properties of gases, liquids, solids, plasmas, etc. An induction based on
the relative entropy $S[p|m]$ may not be of universal validity, but its wide
range of application makes it definitely useful. On the other hand, the
preservation of marginals, which leads to using $\tilde{S}[p|m]$, does not
seem nearly as useful.

It is interesting that if instead of axiomatizing the inference process, one
axiomatizes the entropy itself by specifying those properties expected of a
measure of separation between (possibly unnormalized) distributions one is
led to a continuum of \textquotedblleft entropies,\textquotedblright\ \cite%
{Amari85}\ 
\begin{equation}
S_{\delta }(p|q)=\frac{-1}{\delta (1-\delta )}\int dx\left[ \delta
p+(1-\delta )q-p^{\delta }q^{1-\delta }\right] ~,
\end{equation}%
equivalent, for the purpose of updating, to the relative Renyi entropies 
\cite{Renyi61, Aczel75}. The shortcoming\ of this approach is that it is not
clear when and how such entropies are to be used, which features of a
probability distribution are being updated and which preserved, or even in
what sense do these entropies measure an amount of information. Remarkably,
if one further requires that $S_{\delta }$ be additive over independent
sources of uncertainty, as any self-respecting measure ought to be, then the
continuum in $\delta $ is restricted to just the two values $\delta =0$ and $%
\delta =1$ which correspond to the two entropies derived in this paper: $%
S_{1}$ and $S_{0}$ are equivalent to our $S$ and $\tilde{S}.$ This raises
the interesting question of whether it is possible to identify a $\delta $%
-continuum of alternatives to Axiom 3. To conclude our brief remarks on the
entropies $S_{\delta }$ we point out first, that there exist a variety of
physical examples where it appears that maximizing an $S_{\delta }$ yields
reasonable results (the $S_{\delta }$ are equivalent to the Tsallis'
entropies \cite{Tsallis88}); and second, that there is one very intriguing
suggestion that using $S_{\delta }$ need not be incompatible with a more
standard use of MaxEnt or ME \cite{Plastino94}.

Finally, it is clear that the extended method of maximum entropy which we
have here called ME should allow us to tackle problems which cannot be
envisaged within the more restricted scope of MaxEnt. Two examples are
presented in these proceedings \cite{Tseng03, Caticha03}.

\noindent \textbf{Acknowledgments- }I am very indebted to Carlos Rodr\'{\i}%
guez, Chih-Yuan Tseng, Roland Preuss, Marian Grendar and Alberto Solana for
many insightful remarks and valuable discussions.


\begin{thebibliography}{99}
\bibitem{footnote1} In the original spoken version of this paper I described
the ME method with some emphasis on quantum inference. After hearing, at the
MaxEnt23 Workshop, remarks by P. Cheeseman and L. Bretthorst in which they
correctly point out that MaxEnt, as usually understood, is not a method for
updating probability distributions, I realized that the pedagogical impact
of the written version would be considerably enhanced if I were to change
the emphasis of the presentation in order to bring out the distinction
between ME, which is designed to be a method for inductive inference, and
MaxEnt, which is not. The discussion on quantum relative entropy will be
presented elsewhere.
\bibitem{footnote2} On terminology: The terms `prior' and `posterior' are
normally used in the context of Bayes' theorem; we retain the same
terminology when using ME because we are concerned with the similar goal of
processing information to update from a prior to a posterior.
\end{thebibliography}
\end{document}